\documentclass[reprint,amsmath,amssymb,aps,floatfix]{revtex4-2}
\usepackage{xcolor}
\usepackage{graphicx,epsfig,float,pst-all}
\usepackage{mathrsfs}
\usepackage{inputenc}
\usepackage{textcomp}
\usepackage{mathtools}
\usepackage[normalem]{ulem}
\usepackage{dcolumn}
\usepackage{bm}
\usepackage{hyperref}
\usepackage[title]{appendix}
\begin{document}

\newcommand{\rb}[1]{\textcolor{red}{\it#1}}
\newcommand{\rbout}[1]{\textcolor{red}{\sout{#1}}}

\preprint{APS/123-QED}

\title{Relativistic calculations of molecular electric dipole moments of singly charged aluminium monohalides}

\author{R. Bala}
 \email{balar180@gmail.com}
\author{V. S. Prasannaa}
\email{srinivasaprasannaa@gmail.com}
\author{B. P. Das}
\affiliation{Centre for Quantum Engineering, Research and Education, TCG CREST, Salt Lake, Kolkata 700091, India\\}
\date{\today}

\begin{abstract}
In this work, we have studied the  permanent electric dipole moments of singly charged aluminium monohalides in their electronic ground state, $X^2\Sigma$, using Kramers-restricted relativistic configuration interaction method. We report our results from this method in the singles and doubles approximation with those of Dirac-Fock calculations. %method and the Kramers-restricted configuration interaction method in the singles and doubles approximation. 
For our finite field computations, quadruple zeta basis sets were employed. We discuss the electron correlation trends that we find in our calculated properties and have compared our results with those from literature, wherever available. 
\begin{description}
\item[Keywords]
{\it ab initio} calculations, permanent electric dipole moments 
\end{description}
\end{abstract}

\maketitle

\section{\label{sec:secI}Introduction}

Molecular electric dipole moment is an important property in the field of cold and ultracold molecules, due to its occurrence in a wide range of applications. Molecular electric dipole moments (also known as permanent electric dipole moments, and abbreviated hereafter as PDMs) are useful in the study of long range dipole\,-\,dipole interactions~\cite{Tohme_2016}, which in turn find applications in the fields of quantum phase transitions~\cite{Trefzger_2011, Yu_2007, Lie_2011}, quantum information processing~\cite{Carr_2009}, and quantum computing~\cite{DeMille_2002, Robin_2006}. PDMs are also crucial inputs in searches for physics beyond the Standard Model of elementary particles via the  electric dipole moment of the electron (eEDM)~\cite{Abe_2018, Fazil_2019, Sunaga_2019, Prasannaa_2020}. PDMs also become  significant in studies that strive to improve understanding of electrostatic bonding, for example, see Refs.~\cite{Buckingham_1959}. \\ 

In particular, the PDMs of aluminium monohalide systems may be interesting for many applications in the coming years, given the recent interest in laser cooling of molecular ions and works that have theoretically shown that AlF$^+$ and AlCl$^+$ can be laser cooled~\cite{Kang_2017}. A few studies have been carried out on the two systems prior to identifying that they are amenable to laser cooling~\cite{Dyke_1984,Meyer_1988,Meyer_1991,Brites_2008, Robert_1984, Bruna_2001}. However, to the best of our knowledge, we could not find works that calculated PDMs of the heavier counterparts, AlBr$^+$, AlI$^+$, AlAt$^+$, and the superheavy AlTs$^+$. In this work, keeping in mind the near-future scope of singly ionized molecules and the scarcity of literature for aluminium monofluoride ions, we carry out accurate relativistic calculations of PDMs of these systems with high-quality basis sets. We carry out relativistic calculations as we expect relativistic effects to become increasingly important as we progress from the lighter to the heavier molecules in the series, especially in the case of AlTs$^+$. We also study the correlation effects and their trends in these systems. \\ 

This paper is organized into three sections. The theory and methodology used to calculate the results of the aforementioned molecular properties are given in section~\ref{sec:secII}. Section~\ref{sec:secIII} includes the results and discussion on the computed properties, while section~\ref{sec:secIV} presents a summary of this work. 
%%%%%%%%%%%%%%%%%%%%%%%%%%%%%%%%%%%%
\section{\label{sec:secII}Theory and Methodology}
%%%%%%%%%%%%%%%%%%%%%%%%%%%%%%%%%%%%%
\subsection{Theory}
We have employed the Dirac-Hartree-Fock (DF) as well as Kramers-restricted Configuration Interaction (KRCI) approaches to obtain the PDMs ($\mu$). We use the Dirac-Coulomb Hamiltonian throughout this work. We obtain the property of interest to us using the finite field method (two point central difference) as~\cite{Buckingham_1967} 
\begin{eqnarray}{}
\mu_i\,=\,-\left(\frac{dE}{d\varepsilon_i}\right)_0.
\end{eqnarray} 

In the above expression, $E$ refers to the energy, while $\varepsilon$ is the perturbative parameter. Throughout, we consider the $z$\,-\,axis as the internuclear axis of the considered diatomic molecule. Further details of the approaches employed in this work can be found in Ref.~\cite{Bala_2019}. 
%%%%%%%%%%%%%%%%%%%%%%%%%%%%%%%%%%%%%%%%%%%
\subsection{Details of calculation}
%%%%%%%%%%%%%%%%%%%%%%%%%%%%%%%%%%%%%%%%%%%
We first perform geometry optimization to determine the equilibrium bond lengths of the considered molecules, using relativistic density functional theory (DFT) and with Dyall's v3z basis sets (of triple zeta quality)~\cite{Dyall_2006, Dyall_2016}. The functional used in the DFT calculations is B3LYP. The equilibrium bond lengths are found to be  1.623\,\AA\,for AlF$^+$, 2.058\,\AA\,for AlCl$^+$, 2.223\,\AA\,for AlBr$^+$, 2.512\,\AA\, for AlI$^+$, 2.768\,\AA\,for AlAt$^+$, and 2.928\,\AA\,for AlTs$^+$. All of our computations were carried out in the DIRAC22~\cite{DIRAC} program. \\

The field dependent energies are computed at DF and KRCI levels of theory for all aluminium halides at their optimized values of internuclear distances. For the KRCI calculations, we have considered single\,- and double\,- excitations with respect to the reference DF state (KRCI method in singles and doubles approximation, abbreviated as KRCISD hereafter). In particular, the KRCISD energies are evaluated using generalised active space (GAS) technique. In this approach, the active orbitals are divided into three sub spaces: filled paired (GAS1), filled unpaired (GAS2) and unfilled virtual orbitals (GAS3). We have considered 9 active electrons for all the molecules and imposed an energy cut-off of $10\,E_h$ for the virtuals, in order to avoid the associated steep computational cost, particularly for the heavier systems. \\ 

The reported calculations have been performed using Gaussian charge distribution for nuclei. Further, the aluminium atom is chosen to be at the origin. The perturbative electric field strengths are chosen in the range $-1\times10^{-4}$ to $1\times10^{-4}$ au. Relativistic uncontracted Dyall basis sets of quadruple zeta quality (v4z) have been used for all calculations~\cite{Dyall_2006, Dyall_2016}. The details of the number of basis functions are:

\begin{itemize}
\item Al: ($24s,\,14p,\,3d,\,2f,\,1g$), 
\vspace{-0.3cm}\item F: ($18s,\,10p,\,3d,\,2f,\,1g$), 
\vspace{-0.3cm}\item Cl: ($24s,\,14p,\,3d,\,2f,\,1g$), 
\vspace{-0.3cm}\item Br: ($30s,\,21p,\,13d,\,2f,\,1g$), 
\vspace{-0.3cm}\item I: ($33s,\,27p,\,18d,\,2f,\,1g$), 
\vspace{-0.3cm}\item At: ($34s,\,31p,\,21d,\,14f,\,1g$), and 
\vspace{-0.3cm}\item Ts: ($35s,\,35p,\,24d,\,16f,\,1g$). 
\end{itemize}

We use atomic units throughout this work, unless mentioned otherwise.
%%%%%%%%%%%%%%%%%%%%%%%%%%%%%%%%
\section{\label{sec:secIII}Results and Discussion}
%%%%%%%%%%%%%%%%%%%%%%%%%%%%%%%%%
%%%%%%%%%%%%%%%%%%%%%%%%%%%%%%%%%%%%
\subsection{Equilibrium bond lengths}
%%%%%%%%%%%%%%%%%%%%%%%%%%%%%%%%%%%%
\begin{figure}[t]
%\centering
\addtolength{\tabcolsep}{-0.3cm}
\begin{tabular}{cc}
\includegraphics{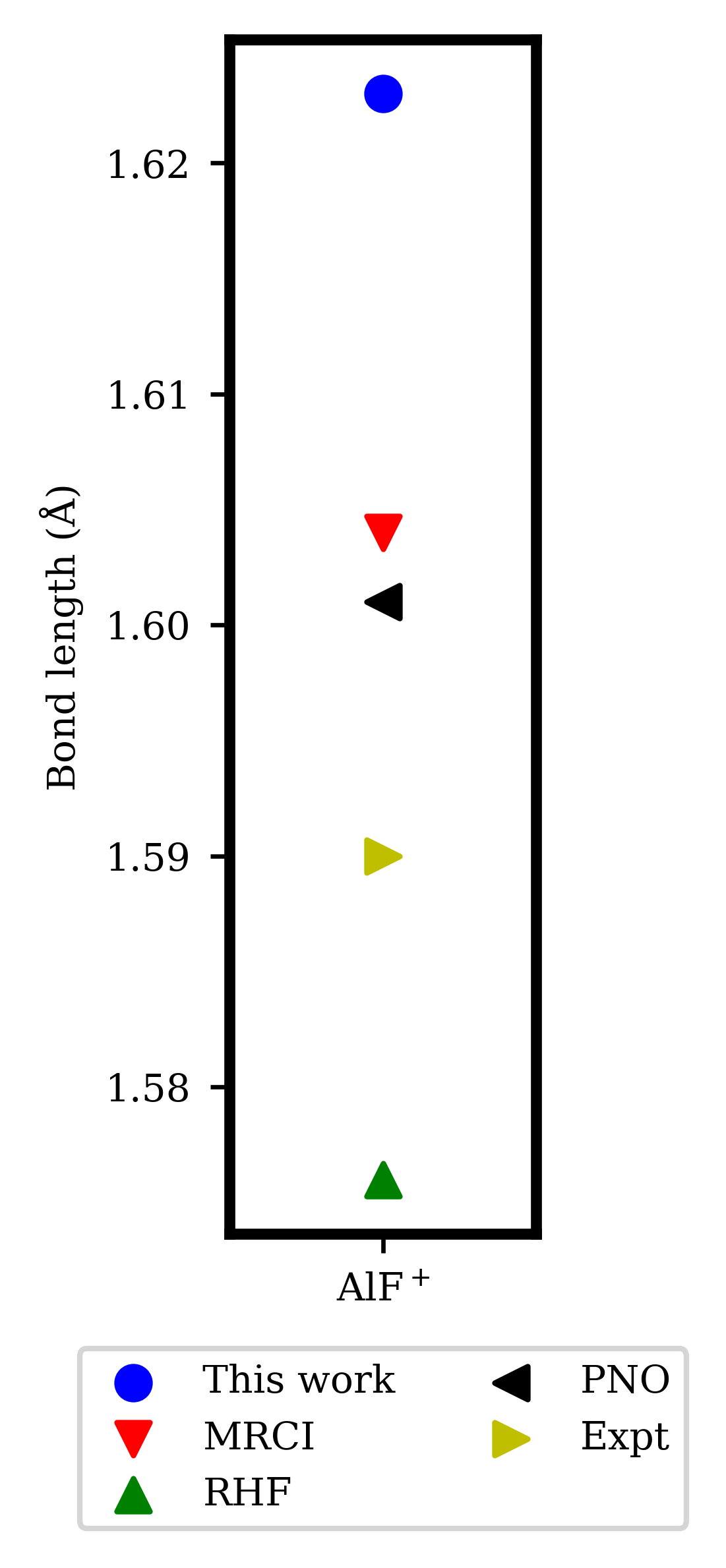}&\includegraphics{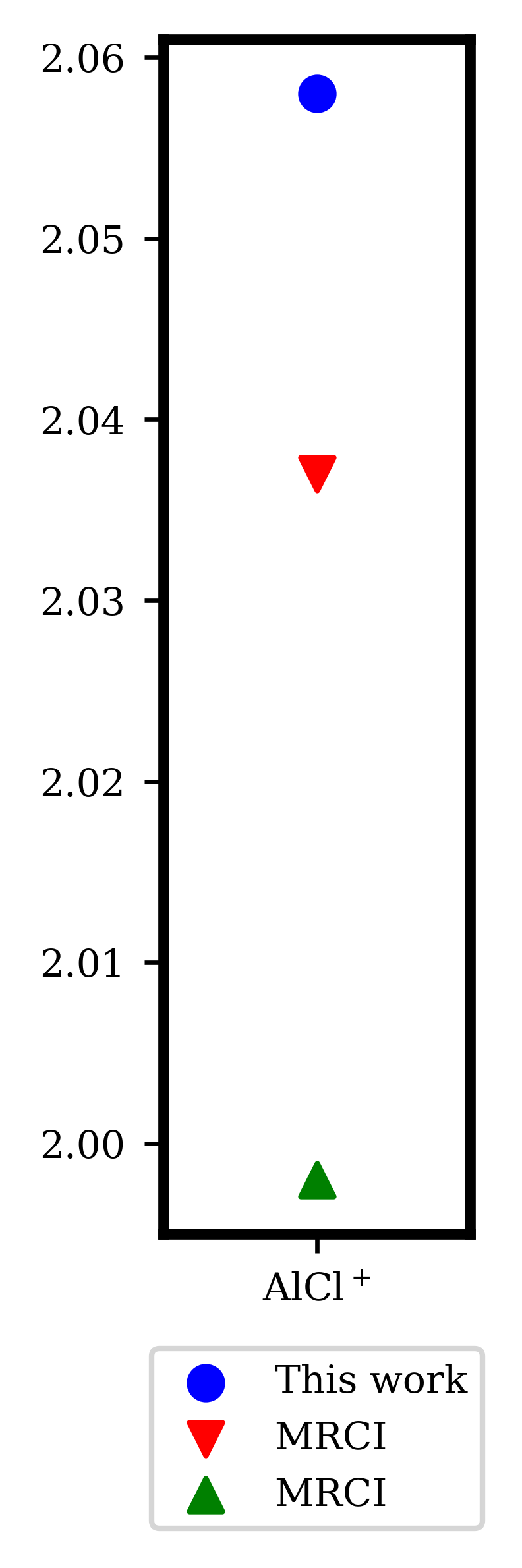}\\
(a)&(b)\\
\end{tabular}
\caption{\label{fig:FIG1} a. Calculated bond lengths (in Angstrom) for AlF$^+$ and b. AlCl$^+$ systems. We have compared it with the available literature values (for AlF$^+$ and AlCl$^+$). In the legends, `MRCI' refers to multireference configuration interaction, `RHF' to restricted Hartree-Fock, `PNO' is pseudo-natural orbitals coupled electron pair approximation, and `Expt' is the abbreviated notation for experiment.}
\end{figure}

\begin{figure}[t]
%\centering
\addtolength{\tabcolsep}{0.1cm}
\begin{tabular}{c}
\includegraphics[width=\columnwidth]{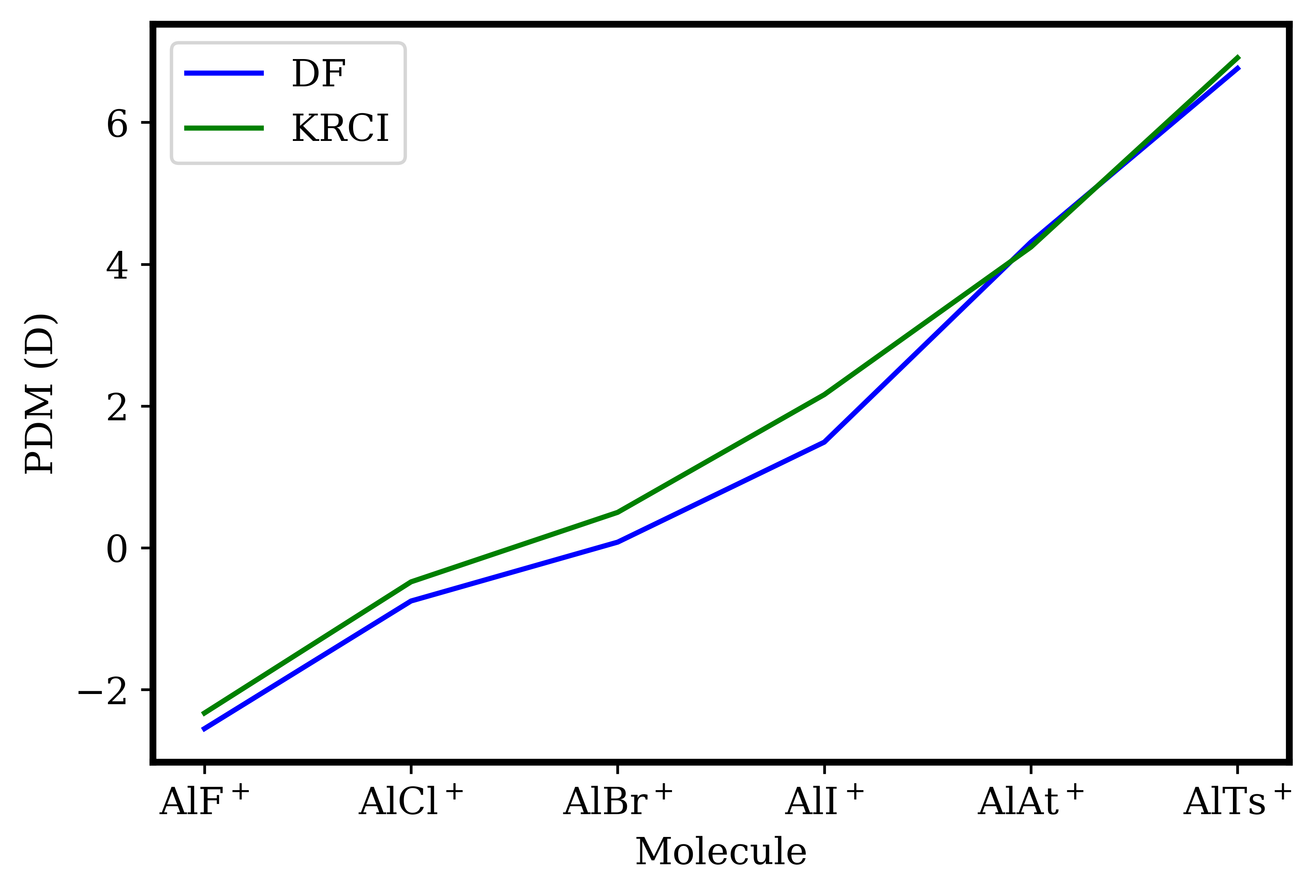}\\
\end{tabular}
\caption{\label{fig:FIG2} Molecular PDMs for the ground states of AlX$^+$ (X: F, Cl, Br, I, and At) ions in Debye (D), at DF (blue) and KRCI (green) levels of theory.}
\end{figure}
From the geometry optimization procedure using relativistic DFT, we find the equilibrium bond lengths to be 1.623, 2.058, 2.223, 2.512, 2.768, and 2.928 \AA\,for AlF$^+$, AlCl$^+$, AlBr$^+$, AlI$^+$, AlAt$^+$, and AlTs$^+$, respectively. The numerical values of all our obtained bond lengths along with literature values (available only for AlF$^+$ and AlCl$^+$) are collected in Table~(\ref{tab:table1}). Figure~(\ref{fig:FIG1}) shows our results and how they compare with the available {\it ab initio} and experimental results for AlF$^+$ and AlCl$^+$. \\ 

The molecular PDMs of aluminium monohalides discussed in the subsequent subsections have been calculated at the corresponding equilibrium bond lengths mentioned above. Our values of equilibrium bond lengths for AlF$^+$ and AlCl$^+$ show good agreement with the existing results. The maximum difference between the values reported in the current work and those from literature using many-body methods is 1.9\% (3.0\%) for AlF$^+$ (AlCl$^+$). Experimental data for equilibrium bond length is available only for AlF$^+$, among all the systems that we considered, and our computed value differs from this result by 2.1\%. Thus, it is reasonable to assume that relativistic DFT with triple zeta quality basis set provides sufficiently accurate results for equilibrium bond lengths of the considered molecules.
%%%%%%%%%%%%%%%%%%%%%%
\subsection{Permanent dipole moments}
%%%%%%%%%%%%%%%%%%%%%%
The magnitude of PDMs for all the considered molecules along with the available data from literature are collected in Table~(\ref{tab:table2}). Figure~(\ref{fig:FIG2}) contains the values of PDMs both at DF as well as KRCISD levels of theory. From the table, it can be seen that with the choice of Al as origin, while AlCl$^+$ and AlBr$^+$ possess the smallest values of PDM, the superheavy AlTs$^+$ has the largest PDM among the aluminium monohalide ions at 6.91 D. It can be seen from the figure that the PDMs increase monotonically as one goes from lighter to heavier systems, with the PDMs for AlF$^+$ and AlCl$^+$ being negative and with those of the heavier systems being positive. The correlation trend itself is found to be oscillatory, with the correlation effects being found to be very large for AlCl$^+$, AlBr$^+$, and AlI$^+$ (about 62, 86, and 31 percent respectively). On the other hand, we find that electron correlation contributes to the PDMs of AlAt$^+$ and AlTs$^+$ by less than 2.2 percent. The comparison of magnitudes of our values of PDMs for AlF$^+$ and AlCl$^+$ with the existing calculations is discussed in the following paragraphs. \\ 

Klein and Rosmus~\cite{Robert_1984} have reported the molecular PDM for AlF$^+$. They have used pseudonatural orbital configuration interaction (PNO CI) method together with the Gaussian type orbitals. The basis sets used in their work include ($12s, 9p, 2d$) and ($10s, 6p, 2d$)  functions for Al and F atoms, respectively. Further, the deepest five $s$ and four $p$ functions for Al and four $s$ and three $p$ functions for F have been used in their contracted form. At the CI level, they have considered single and double exciations with respect to the reference Hartree-Fock (HF) state. To evaluate the effects of excitations higher than single and double excitations, the coupled electron pair approximation (CEPA) has been utilized in their work. The PNO CISD method and CEPA approximation together is abbreviated as PNO\,-\,CEPA. Since they have computed their PDM values in centre-of-mass system, we have translated the PDM by considering Al atom as origin. With this transformation, their PDM for AlF$^+$ ionic system is 2.29\,D. Our PDM value of 2.56\,D at DF level is larger than the HF value given in their work by 7.1\%. The correlated value of PDM in Ref.~\cite{Robert_1984} differs from our value at KRCISD level by approximately 1.7\%. \\  

Kang {\it et al.}~\cite{Kang_2017} have computed the PDM for AlCl$^+$ molecule using the MRCI method with Davidson correction (MRCI+Q). They  obtain a PDM of $\approx$ 0.27\,D, which is smaller by 0.2 D with respect to our work (0.47 D). We note that we use the KRCI approach, while they employ MRCI method. Besides this difference, we consider the effects of relativity, bond length, choice of active space, and choice of basis, all using KRCI, to possibly explain the dependence of PDM on these factors.\\ 

\emph{Relativity-- }The authors in Ref.~\cite{Kang_2017} had incorporated spin-orbit coupling effects in their calculations. We indirectly checked for its significance by calculating the PDM using 4-component spinfree (0.43 D) as well as non-relativistic (0.45\,D) approaches, with the former including  mass-velocity-Darwin effects. Noting that our calculation with the Dirac-Coulomb Hamiltonian had yielded 0.47\,D, we expect that relativity  contributes negligibly to the PDM. \\ 

\emph{Bond length-- }Further, the difference in bond length between our work and theirs is 0.06 \AA. We carried out KRCI calculations with the bond length used in Kang {\it et al.}~\cite{Kang_2017} (= 1.998 \AA) and with 10 $E_h$ cut-off for virtuals, and found the PDM to be 0.30 D (versus Kang {\it et al.}'s 0.27 D). Thus, difference in bond length between our work and theirs was one source of discrepancy. \\ 

\emph{Active space and single-particle basis-- }We note that their active space is much smaller than ours. They had considered 9 active electrons and 3 virtual molecular orbitals. On the other hand, our active space for relativistic correlation calculation includes 9 active electrons and 115 virtual orbitals. Moreover, Kang {\it et al.}~\cite{Kang_2017} employ the aug-cc-pV5Z basis, while we employ Dyall's quadruple zeta quality basis sets. It is worth noting that within a given virtual cut-off energy, the number of virtuals can vary between the two bases. Therefore, we carried out KRCI calculations (all with 9 active electrons) taking these possibilities into account, which we summarize below: \\ 

\begin{itemize}
\item Our main result: Dyall v4z basis, 115 virtual orbitals (10 $E_h$ cut-off) gives 0.47 D.
\item Dyall v4z basis, 43 virtual orbitals (1 $E_h$ cut-off) gives 0.17 D. 
\item cc-pVQZ basis, 111 virtual orbitals (10 
$E_h$ cut-off) gives 0.47 D. 
\item cc-pVQZ basis, 37 virtual orbitals (1 $E_h$ cut-off) gives 0.15 D. 
\end{itemize}

We emphasize at this point that a space of 3 virtual orbitals would correspond to a cut-off much lesser than 1 $E_h$, and we carried out calculations with 1 $E_h$ cut-off due to practical limitations with the employed code. A comparison of results 1 and 2, as well as 3 and 4 indicates that reduction in important low-lying virtual orbitals can alter the result, while comparing results 1 and 3, and 2 and 4, point at the result not being very sensitive to choice of basis (in this case Dyall v4z vs cc-pVQZ).\\ 

We now compare our result with that obtained in Ref.~\cite{Meyer_1991}. The authors employ MRCI, and obtain a PDM of 0.19 D. The size of their basis for both the atoms is ($17s,\,12p)\,\rightarrow\,[6s,\,5p$], and is augmented with three $d$ and one $f$ functions. Our basis size is substantially larger. Their choice of active space is also significantly smaller than ours. Further, we do not expect much difference from choice of bond lengths, as they chose 2.037 \AA, which is not very different from our 2.058 \AA. \\ 

We now consider the sources of error in our calculations. We consider AlF$^+$ as a representative system for the error budget, and consider the choice of basis (quality and role of diffuse functions), excitations (in particular, triples), number of active electrons, the number of virtual orbitals, and finally the choice of bond length. All these computations have been done at the double zeta level, in view of the involved computational cost with a higher quality basis. We expect that the error budget estimated with the double zeta basis would not be very different from that for a higher quality basis. We have found the error due to basis set size is at most -2.1 percent. We estimated this value based on our expectation that the difference in PDM between a pentuple zeta basis and the quadruple zeta one would not exceed that between the quadruple zeta and a triple zeta basis. Further, the inclusion of augmented functions to the basis was found to change the PDM by -5.1 percent. The difference in the value of PDM  obtained at the equilibrium bond length reported in the current work and that obtained at the experimental value for AlF$^+$ is found to be -3.8 percent. This accounts for the possible error due to difference in bond length. The inclusion of the triple excitations along with singles and doubles changes the PDM by -3.8 percent. To understand the effect of high lying virtuals, we have performed the PDM calculation with 15 $E_h$ virtual cut-off. The PDM with 15 $E_h$ virtual cut-off is found to be 2.49 D that contributes to a change of 0.4 percent. Further, the inclusion of 15 electrons instead of 9 electrons in the active space changes the PDM by 1.2 percent. Finally, we also considered the effect of redistributing active orbitals between GAS1 and GAS2, and found that when we flip the distribution, that is, 1 orbital in GAS1 and 4 in GAS2 in place of 4 in GAS1 and 1 in GAS2, the results change by -0.8 percent. \\ 

By considering all the factors discussed above, we expect the maximum error in our computed results to be about 12 percent. \\ 
%%%%%%%%%%%%%%%%%%%%%%%%%%%%%%%%%%%%
\section{\label{sec:secIV}Summary}
%%%%%%%%%%%%%%%%%%%%%%%%%%%%%%%%%%%%%%%
In summary, we report results for equilibrium bond lengths and the PDMs of aluminium monohalides using high quality basis sets and accurate many-body theory, in view of scarcity in literature for the heavier systems beyond AlF$^+$ and AlCl$^+$. Our findings of equilibrium bond lengths and PDMs for the two lighter systems match well with the available results. We also report the trends in PDMs across the considered systems. We find that correlation effects can be very significant, for example, in AlCl$^+$, AlBr$^+$, and AlI$^+$. In order to understand the difference in the PDM of AlCl$^+$ between our result and a literature value, we carry out a detailed analysis to account for the effects of relativity, bond length, choice of active space, and basis sets. We find that for AlCl$^+$, the size of the virtual orbital space as well as relatively small changes in equilibrium bond length could affect the results significantly, while relativistic effects may not play such a vital role. Finally, we also choose AlF$^+$ as a representative system for estimating the error budget, and find that the net effect due to errors on the PDM is about 12 percent. The accuracy of the results of the PDM calculations in the present work can be assessed when the PDMs of these laser coolable molecular ions become available in the future by accurate spectroscopic measurements.

\begin{acknowledgments}
All calculations reported in the current work were carried out on the Rudra cluster, Sankhyasutra Labs (Bangalore, India). 
\end{acknowledgments}

\bibliography{AlX+_ions}% Produces the bibliography via BibTeX.

\newpage
\onecolumngrid
\appendix

\section*{Appendix}
The numerical values of all the computed quantities in this work along with the existing results are given in appendix. 

\begin{table}[htbp]
\setcounter{table}{0}
\renewcommand{\thetable}{A\arabic{table}}
\caption{\label{tab:table1}Equilibrium bond lengths (in \AA) for AlX$^+$ (X: F, Cl, Br, I , At, and Ts) molecular systems using relativistic DFT with triple zeta basis sets. We also compare our obtained values with those available in literature. }
\begin{ruledtabular}
\begin{tabular}{lllllll}
 Molecule & Method &  $R_e$\,(\AA) & Ref.\\
\hline
AlF$^+$  & DFT & 1.623 & This work\\
         & MRCI & 1.593 & \cite{Kang_2017}\\
         & RHF   & 1.576 & \cite{Dyke_1984}\\
         & PNO-CEPA & 1.601 & \cite{Dyke_1984}\\
         & MRCI & 1.604 & \cite{Meyer_1991}\\
         & Expt & 1.59\,$\pm$\,0.01 & \cite{Dyke_1984}\\
AlCl$^+$ & DFT & 2.058 & This work\\
         & MRCI & 1.998 & \cite{Kang_2017}\\
         & MRCI & 2.037 & \cite{Meyer_1991} \\
         & MRCI & 2.023 & \cite{Brites_2008}\\
AlBr$^+$ & DFT & 2.223 & This work\\
AlI$^+$  & DFT & 2.512 & This work\\
AlAt$^+$ & DFT & 2.768 & This work\\
AlTs$^+$ & DFT & 2.928 & This work\\
\end{tabular}
\begin{flushleft}
\end{flushleft}
\end{ruledtabular}
\end{table}

\begin{table}[htbp]
\setcounter{table}{1}
\renewcommand{\thetable}{A\arabic{table}}
\caption{\label{tab:table2}Magnitude of permanent dipole moments (in Debye) for AlX$^+$ (X: F, Cl, Br, I, At, and Ts) molecular systems at DF and KRCISD level using quadruple zeta (dyall.v4z) basis sets.}
\begin{ruledtabular}
\begin{tabular}{llllllll}
 Molecule & Method & PDM$^\#$ & Ref.\\
\hline
AlF$^+$  & DF     & 2.56 & This work \\
         & KRCISD   & 2.33 & This work \\
         & MRCI$^\dag$  & 2.40 & \cite{Meyer_1991}\\
         & HF$^\dag$  & 2.39 & \cite{Robert_1984}\\
         & PNO-CEPA$^\dag$ & 2.29 & \cite{Robert_1984}\\
\hline
AlCl$^+$ & DF      & 0.76 & This work \\
         & KRCISD   & 0.47 & This work \\
         & MRCI$^{\dag,a}$ & 0.27 & \cite{Kang_2017}\\
         & MRCI$^\dag$     & 0.19 & \cite{Meyer_1991}\\
\hline
AlBr$^+$ & DF      & 0.07 &  This work\\
         & KRCISD  & 0.50 & This work\\
\hline
AlI$^+$  & DF      & 1.49 & This work\\
         & KRCISD   & 2.16 & This work\\
\hline
AlAt$^+$ & DF     & 4.31 & This work\\
         & KRCISD & 4.24 & This work\\
\hline
AlTs$^+$ & DF & 6.76 & This work\\
         & KRCISD & 6.91 & This work\\
\end{tabular}
\begin{flushleft}
$^\#$Results for PDM are rounded off at second decimal place.\\
$^\dag$In Ref.~\cite{Meyer_1991, Robert_1984}, the authors have reported the values of PDM  in centre-of-mass system. We have suitably translated them with Al chosen to be the origin to make the comparison consistent with our values. \\
$^a$The PDM has been extracted from the graph given in Ref.~\cite{Kang_2017}.\\
\end{flushleft}
\end{ruledtabular}
\end{table}

\end{document}